\documentstyle[amsfonts,12pt,thmsa,sw20lart]{article}
\input{tcilatex}
\begin{document}

\author{R.\ Peschanski\thanks{email: pesch@spht.saclay.cea.fr}}
\title{
Dual Shapiro-Virasoro amplitudes in the QCD dipole picture
}
\maketitle
\begin{center}
Service de Physique Th\'{e}orique

CEA-Saclay

F-91191 Gif-sur-Yvette

FRANCE
\end{center}

\begin{abstract}

Using the QCD dipole picture of BFKL dynamics and the conformal invariance properties of the BFKL kernel in transverse coordinate space, we show that the $%
1\!\rightarrow \!p$ dipole densities can be expressed in terms of dual
Shapiro-Virasoro amplitudes $B_{2p+2}$ and their generalization including non-zero conformal spins.\ We  discuss  the possibility of an effective closed string theory  of interacting QCD dipoles.

\end{abstract}
\newpage

\section{\protect\medskip BFKL equation and  QCD dipole picture
}

The  QCD ``hard Pomeron''  is understood as  the solution of perturbative QCD expansion at high energy ($W$) after resumming 
the leading $\left( \alpha \ln W^{2}\right)^n$ terms. It is known to obey 
the BFKL equation\cite{BFKL}. It has
recently attracted a lot of interest in relation with the experimental
results obtained at HERA for deep-inelastic scattering reactions at very low
value of $x_{Bj}\approx Q^{2}/W^{2},$ where $Q^{2}$ is the virtuality of the
photon probe $\gamma ^{*}$ and $W$ is, in this case, the c.o.m energy of the $\gamma ^{*}$%
-proton system.\ Interestingly enough, the proton structure functions
increase with $W$ at fixed $Q$\cite{DESY} in a way qualitatively compatible with the  prediction of the BFKL equation. However, the phenomenological discussion is still under way,
 since scattering of a ``hard'' probe on a proton is not a fully perturbative QCD process and moreover, alternative explanations based on renormalization group evolution equations do exist\cite{DESY}. On the other hand, the phenomenological success\cite
{Proton} of parametrizations based on the BFKL evolution in the framework of the
QCD dipole model\cite{Muller} is quite encouraging for a further study of its properties.

Beyond these  phenomenological motivations, there exist interesting related
theoretical problems which we want to address in the present paper.\ In its
 2-dimensional version, the BFKL equation expresses\cite{BFKL} the leading-order resummation
result for  the elastic (off-mass-shell) gluon-gluon scattering amplitude in the
$2\!-\!d$ transverse plane  $f\left(
k_{0}k_{1};k_{0}^{\prime }k_{1}^{\prime }|Y\right) ,$ which depends  on the energy $W=e^{Y/2}$
and on the 2-momenta of the incoming and outgoing gluons with $q=k_{0}^{\prime
}-k_{0}=k_{1}^{\prime }-k_{1}$ is the transferred 2-momentum. $Y$
is the total rapidity space available for the gluon-gluon reaction. Alternatively, one introduces\cite
{Lipatov,Muller}  the coordinate variables via
2-dimensional Fourier transforms and the gluon Green function $
f\left(\rho _{0}\rho _{1};\rho _{0}^{\prime }\rho _{1}^{\prime }|Y\right) $
where 2-momentum conservation in momentum space  leads to  global
translationnal invariance in coordinate space. The amplitude $f\left( \rho _{0}\rho _{1};\rho _{0}^{\prime }\rho
_{1}^{\prime }|Y\right) $ is solution of the BFKL equation expressed in the 
2-dimensional transverse coordinate space and explicit solutions can be obtained\cite
{Lipatov,Navelet} using conformal invariance properties of the BFKL kernel. We will
 heavily use these  symmetry properties in the sequel.

In our paper, we will address the problem of finding the amplitudes $f^{\left( p\right) }$ solution of
  processes
involving $2p+2$ external gluon legs where $\rho _{0}\rho
_{1}, $ $\rho _{a_0}\rho _{a_1},$...,$\rho _{p_0}\rho _{p_1}$ are their arbitrary
coordinates in the plane transverse to the initial gluon-gluon direction. Note
that $f^{\left( 1\right) }\equiv $ $ f\left(\rho _{0}\rho _{1};\rho _{0}^{\prime }\rho _{1}^{\prime }|Y\right)$ is  the original BFKL amplitude. Recently,
 it
has been shown\cite {Navelet} that $f^{\left( 1\right) }$ is equal, up to kinematical factors, to  the
number density $n_{1}$ of dipoles existing in the wave-function of an
initial quark-antiquark pair (onium) after an evolution ``time''$Y$ (such can be interpreted $Y$ in the BFKL equation written as a diffusion process\cite {bartels}). In the QCD-dipole picture\cite {Muller}%
, gluons are equivalent to a $q \bar q$ pairs (in the $N_{c}\!\rightarrow\!
\infty $ limit) which recombine into a collection of independent and colorless
dipoles. The elastic amplitude is thus obtained from the elementary
dipole-dipole amplitude weighted by the dipole number densities of each
initial state obtained after evolution time $Y.$ Using conformal invariance properties of the BFKL
kernel, it turns out\cite{Navelet} that:
\begin{equation}
n_{1}^{n,\nu }\left( \rho _{0}\rho _{1};\rho _{0}^{\prime }\rho _{1}^{\prime
}\right) ={\scriptstyle \frac{2\ \left(\nu ^{2}+n^{2}/4\right)}{\pi ^{4}\left| \rho _{0^{\prime
}1^{\prime }}\right| ^{2}}}\ f^{\left(n,\nu \right) }\left( \rho _{0}\rho
_{1};\rho _{0}^{\prime }\rho _{1}^{\prime }\right)  \tag{1}
\end{equation}
\noindent where $n_{1}^{n,\nu } $(resp. $f^{\left(n,\nu \right) })$ are the
components of the dipole density (resp. the gluon Green's function) expanded upon the conformally invariant
basis, namely:
\begin{eqnarray}
n_{1}\left( \rho _{0}\rho _{1};\rho _{0}^{\prime }\rho _{1}^{\prime
}|Y\right) &=&\int d\omega \ e^{\omega Y}\sum_{n\in \Bbb Z} \int d\nu \ \frac{n_{1}^{n,\nu }\left( \rho _{0}\rho _{1};\rho
_{0}^{\prime }\rho _{1}^{\prime }\right) }{\omega -\omega \left(n,\nu
\right) }
\nonumber \\
&=&
\sum_{n\in \Bbb Z} \int d\nu \ e^{-\omega \left(n,\nu
\right) Y}\ n_{1}^{n,\nu }\left( \rho _{0}\rho _{1};\rho
_{0}^{\prime }\rho _{1}^{\prime }\right),  \tag{2}
\end{eqnarray}
\noindent and
\begin{equation}
\omega \left( n,\nu \right) =\frac{2\alpha N_{c}}{\pi}\func{Re}\left\{ \psi
\left( 1\right) -\frac{1}{2}\psi \left( 1/2 (1+n)+i\nu \right) \right\} ,  \tag{3}
\end{equation}
\noindent is the value of the BFKL kernel in the (diagonal) conformal basis.
The corresponding eigenvectors are explicitly known\cite{Lipatov} to be:
\begin{equation}
E^{n,\nu }\left( \rho _{o\gamma },\rho _{1\gamma }\right) =\left( -1\right)
^{n}\left( \frac{\rho _{0\gamma }\rho _{1\gamma }}{\rho _{01}}\right)
^{\Delta }\left( \frac{\bar{\rho} _{0\gamma }\bar\rho _{1\gamma }}{%
\bar{\rho} _{01}}\right)^{\widetilde{\Delta }},  \tag{4}
\end{equation}
\noindent with $\rho _{ij}=\rho _{i}\!-\!\rho _{j} \left({\rm  resp.}\ \bar{\rho}
_{ij}=\bar{\rho _{i}}\!-\!\bar{\rho _{j}}\right) $ are the
holomorphic (resp. antiholomorphic) components in the 2-d transverse plane
considered as $\cal C$ and $\Delta =n/2+1/2-i\nu$, $ \widetilde{\Delta }%
=-n/2+1/2-i\nu$ ($n\!\in \! {\Bbb Z}$, ${\nu }\!\in  \!{\Bbb R}$), are the
quantum numbers defining the appropriate unitary representations\cite{Lipatov} of the conformal group $SL(2,{\Bbb C})$. Indeed the BFKL solution (i.e.,
also, the QCD dipole solution\footnote{ There is a controversy\cite {mullercheng} about the equivalence between the BFKL and the QCD dipole approach concerning a discrepancy in the separate evaluation of real and virtual corrections. In the processes we consider, both virtual and real contributions are included in the BKFL 
kernel  and the conformal symmetry of the QCD dipole model calculations  preserves this compensation in the calculations\cite {Navelet}.}) is given by 
\begin{equation}
f^{n,\nu}\left( \rho _{0}\rho _{1};\rho _{0}^{\prime }\rho
_{1}^{\prime }\right) =\int_{{\Bbb R}^{2}}d^{2}\rho _{\gamma }\ \bar{E}
^{n,\nu}\left( \rho _{0\gamma }^{\prime },\rho _{1\gamma }^{\prime }\right)
E^{n,\nu}\left( \rho _{0\gamma ,}\rho _{1\gamma }\right) ,  \tag{6}
\end{equation}
which can be explicitly calculated in terms of hypergeometric functions\cite
{L.N.}.

In order to generalize these investigations to an arbitrary number of
gluons, we shall use the QCD dipole formalism allowing to express the probability of finding $p$ dipoles in an initial one, i.e. the 
p-uple dipole density after an evolution ''time'' $Y,\ $ 
$n_p\left( \left. \rho _{0}\rho _{1};\rho _{a_0}\rho _{a_1},\rho _{b_0}\rho
_{b_1},...,\rho _{p_0}\rho _{p_1}\right| Y\right) . 
$ $n_p$ is the solution of an integral equation which has been proposed in the paper of  Ref.\cite{A. Muller},
and approximate solutions have been worked out and applied to problems like the
double and triple-QCD Pomeron coupling\cite{A. Muller}, the QCD dipole production\cite{Bialas1}, hard diffraction\cite{Bialas} and, more
generally, to the unitarization problem\cite{A. Muller}. In particular, Monte-Carlo
simulations of the unitarization series based on a numerical resolution of the 
$n_p$ integral equations have been performed\cite{muller}. However a general expression for the solution of
these equations and a physical interpretation of its properties are yet lacking. It is the purpose of our work to provide
such a solution, which is intimately related, as we shall see, to dual 
string amplitudes emerging from the QCD dipole picture.

Our main result is the explicit expression of the p-uple dipole density distributions in the transverse coordinate plane as dual Virasoro-Shapiro
amplitudes\cite {Shapiro-Virasoro,Frampton} (for conformal spins all equal to $0).$ These are the dominant contributions at high $Y.$
We also give the expressions for arbitrary conformal spins (i.e. for all the  conformal components) in terms of a well-defined generalization of  Shapiro-Virasoro amplitudes. 

The paper is organized as follows:
In section {\bf 2}, we derive the  QCD dipole equation for $n_2$ (for zero conformal spins). The solution is found in a compact form in terms of integrals over explicit conformal eigenvectors. In section {\bf 3} we reformulate the obtained 3-dipole vertex in terms of the Koba-Nielsen projective-invariant parametrization of the Shapiro-Virasoro amplitude $B_6.$ In the following section {\bf 4} we show that the solution  can be iterated and obtain  the triple dipole density distribution in terms of $B_8.$ This iteration gives  the p-uple dipole distribution in terms of  $B_{2p+2}$ integrands. In section {\bf 5} we discuss the extension for arbitrary conformal spin and  the possibility of finding a target-space realization of the 
underlying closed string picture emerging from the QCD dipole interactions.

\section{\protect\medskip Solution of the dipole equation for $n_{2}$}

The integral equation satisfied by the dipole pair density $n_{2}, $ see Fig.1, is
written as follows\cite{A. Muller} 
\begin{eqnarray}
&&n_{2}\left( \left. \rho _{0}\rho _{2};\rho _{a_0}\rho _{a_1},\rho
_{b_0}\rho _{b_1}\right| Y\right) =\frac{\alpha N_c}{\pi }\int_{{\Bbb R}^{2}}\left| \frac{\rho
_{01}}{\rho _{12}\rho _{02}}\right| ^{2}d^{2}\!\rho _{2}\int_{0}^{Y}dy\;
\nonumber \\
&&e^{-%
\frac{2\alpha Nc}{\pi }\left( Y-y\right) \ell n\left| \frac{\rho _{10}}{\epsilon}%
\right| }   \times \  n_{1}^{}\left( \left. \rho _{0}\rho _{2};\rho _{a_0}\rho _{a_1}\right|
y\right)\ n_{1}\left( \left. \rho _{1}\rho _{2};\rho _{b_0}\rho _{b_1}\right|
y\right) \nonumber \\
&&+\ \frac{\alpha N_c}{\pi ^{2}}\dint\limits_{\left| \frac{\rho
_{20}}{\epsilon}\right| ,\left| \frac{\rho _{21}}{\epsilon}\right| >{}1}\left| \frac{\rho
_{01}}{\rho _{12}\rho _{02}}\right| ^{2}d^{2}\rho _{2}\int_{0}^{Y}dy\;e^{-%
\frac{2\alpha Nc}{\pi }\left( Y-y\right) \ell n\left| \frac{\rho _{10}}{\epsilon}%
\right| }\times  \nonumber \\
&&\ \ \ \ \ \times \frac 12 \left\{  n_{2}\left( \left. \rho _{0}\rho _{2};\rho _{a_0}\rho _{a_1},\rho
_{b_0}\rho _{b_1}\right| y\right)\ +\ \left(\rho_0 \iff \rho_1\right)\right\} ,  \tag{7}
\end{eqnarray}
where the integration domain $\left| \frac{\rho
_{20}}{\epsilon}\right| ,\left| \frac{\rho _{21}}{\epsilon}\right| >{}1$ avoids the singularity at the
initial dipole end-points $\rho_0,\rho_1.$ The physical meaning of this equation is
transparent: The probability of finding a pair of dipoles at
given transverse coordinates at time $Y$ is
made of two terms ; The
last term in (7) is the survival probability after $Y$ of two 
dipoles, whereas the former corresponds to the  probability of
creating two new dipoles $\left( \rho _{0}\rho _{2}\right) $ and $\left( \rho
_{1}\rho _{2}\right) $ which then are surviving till $Y$, each with probability $n_{1}$. In both cases, the survival probability is given by the $\epsilon$%
-dependent Sudakov-like form factor $e^{-\frac{2\alpha N_c}{\pi }\left( Y-y\right)
\ell n\left| \frac{\rho _{10}}{\epsilon}\right| }.$

In fact, it is possible\cite{Bialas1} to reexpress this equation in a way which is explicitely independent of
 $\epsilon $ when $\epsilon\!\rightarrow\! 0.$
Multiplying both terms of Eqn.(7) by $e^{\frac{2\alpha N_c}{\pi }\ {\ell n\left| 
\frac{\rho _{10}}{\epsilon}\right| Y}}$ and differentiating with respect to $Y$%
, one gets: 
\begin{eqnarray}
&&\frac{\partial n_{2}}{\partial Y}\left( \rho _{0}\rho _{1};\rho _{a_0}\rho
_{a_1},\rho _{b_0}\rho _{b_1} | Y\right)  \nonumber \\
&=&\frac{\alpha N_c}{\pi }\stackunder{{\Bbb R}^{2}}{\int }\left| \frac{\rho
_{01}}{\rho _{12}\rho _{02}}\right| ^{2}d^{2}\rho _{2}\ n_{1}\left( \rho
_{0}\rho _{2};\rho _{a_0}\rho _{a_1}|Y\right) n_{1}\left( \rho _{1}\rho
_{2};\rho _{b_0}\rho _{b_1}|Y\right) +  \nonumber \\
&+&\frac{\alpha N_c}{2\pi ^{2}}\stackunder{{\Bbb R}^{2}}{\int }\left\{{\cal
L}ip\right\} d^{2}\rho _{2}\  \left\{  n_{2}\left( \left. \rho _{0}\rho _{2};\rho _{a_0}\rho _{a_1},\rho
_{b_0}\rho _{b_1}\right| Y\right)\ +\ \left(\rho_0 \iff \rho_1\right)\right\} 
,  \tag{8}
\end{eqnarray}
where $\left\{{\cal L}ip\right\} $ denotes a parametrization\cite{A. Muller} of the
BFKL kernel in coordinate space, namely 
\begin{equation}
\left\{ {\cal L}ip\right\} =\left| \frac{\rho _{01}}{\rho _{12}\rho _{02}}\right|
^{2}-\ \frac{1}{2\pi }\ln \left| \frac{\rho _{01}}{\epsilon}\right| \left\{ \delta
^{\left( 2\right) }\left( \rho _{12}\right) +\delta ^{\left(
2\right) }\left( \rho _{02}\right) \right\},  \tag{9}
\end{equation}
\noindent with the property\cite {Muller,A. Muller}
$$\stackunder{{\Bbb R}^{2}}{\int }\left\{{\cal
L}ip\right\} d^{2}\rho _{2}\ E^{n,\nu}\left( \rho _{0\gamma },\rho _{2\gamma }\right)\ =\ \omega \left( n,\nu \right)\ E^{n,\nu}\left( \rho _{0\gamma },\rho _{1\gamma }\right). $$
In equation (8), the $\epsilon \!\rightarrow \!0$ limit can be
 taken without harm, thanks to the non-singular behaviour of the kernel $\left\{{\cal L}ip\right\} $ after  regularization. The
kernel $\left\{{\cal L}ip\right\} $ is conformally invariant\cite {Lipatov} and its eigenvalues
and eigenfunctions are given by equations (3) and (4), respectively.

In order to solve Eqn.(8), we shall expand the function $n_{2}$ on the
 basis (4) by writing: 
\begin{eqnarray}
n_{2}\!\!\!\!\!\!\!\!&&\left( \left. \rho _{0}\rho _{1};\rho _{a_0}\rho _{a_1},\rho _{b_0}\rho
_{b_1}\right| Y \right) =\stackunder{i{\Bbb R}}{\int }d\omega \ e^{\omega Y}\stackunder{n\in {\Bbb Z}}{%
\sum }\stackunder{\Bbb R}{\int }d\nu\stackunder{{\Bbb R}^{2}}{\int }d^{2}\rho
_{\gamma }\times  \nonumber \\
&&\ \ \ \times \ E^{n,\nu}\left( \rho _{0\gamma },\rho _{1\gamma }\right)
\ n_{2}^{n,\nu}\left( \left. \rho _{\gamma };\rho _{a_0}\rho _{a_1},\rho _{b_0}\rho
_{b_1}\right| \omega \right) ,  \tag{10}
\end{eqnarray}
where $\left( \omega ,Y\right) $ are conjugate variables and $\rho
_{\gamma }$ is an auxiliary transverse coordinate variable which labels the
set of eigenfunctions $\bar E^{n,\nu}.$ The solution of Eqn.(8) is obtained
 using  the known\cite{Lipatov} orthogonality
relations of the $E^{n,\nu}$ eigenfunctions 
\[
\int \frac{d^{2}\rho _{0}d^{2}\rho _{1}}{\left| \rho _{01}\right| ^{4}}%
\ E^{n,\nu} {\left( \rho _{0\gamma },\rho _{1\gamma }\right) }\bar{E}^{n^{\prime
},\nu
^{\prime }} {\left(
\rho _{0\gamma ^{\prime }},\rho _{1\gamma ^{\prime }}\right) }= 
\]
\begin{equation}
=a_{n,\nu}\delta _{n,n^{\prime }}\delta _{\left( \nu-\nu^{\prime }\right) }\delta
^{\left( 2\right) }\left( \rho _{\gamma \gamma ^{\prime }}\right) +\left(
-1\right) ^{n}b_{n,\nu}\left| \rho _{\gamma \gamma ^{\prime }}\right| \left( 
\frac{\rho _{\gamma \gamma ^{\prime }}}{\bar \rho _{\gamma \gamma
^{\prime }}}\right)^n \delta _{n,-n^{\prime }}\delta \left( \nu+\nu^{\prime
}\right) ,  \tag{11}
\end{equation}
with  $b_{n,\nu}$ given in ref.(5) and
\begin{equation}
a_{n,\nu}\equiv  \frac{\left|b_{n,\nu}\right| ^{2}}{2\pi ^{2}}=\frac{\pi ^{4}/2}{%
\nu^{2}+n^{2}/4} \ .  \tag{12}
\end{equation}
Inserting the decomposition (10) into Eqn.(3), we integrate both sides of the equation by
\begin{equation}
\int dY\ e^{-\omega Y}\ \frac{d^{2}\rho _{0}d^{2}\rho _{1}}{\left| \rho
_{01}\right| ^{4}}\ \bar{E}^{n,\nu}\left( \rho _{0\gamma },\rho _{1\gamma
}\right)  \tag{13}
\end{equation}
one finds using relations (11) 
\begin{eqnarray}
n_{2}^{n,\nu}\left( \left. \rho _{\gamma };\rho _{a_0}\rho _{a_1},\rho
_{b_0}\rho _{b_1}\right| \omega \right) =\frac{1}{2a_{n,\nu}\left( \omega -\omega (n,\nu)\right) }\times  \nonumber \\
\times \frac{\alpha N_c}{\pi ^{2}}\int \frac{d^{2}\rho _{0}d^{2}\rho
_{1}d^{2}\rho _{2}}{\left| \rho _{01}\ \rho _{02}\ \rho _{12}\right| ^{2}}%
\ E^{n,\nu}{\left( \rho _{0\gamma },\rho _{1\gamma }\right) }\int dY\ e^{-\omega
Y}  \nonumber \\
\times n_{1}\left( \left. \rho _{0}\rho _{1},\rho _{a_0}\rho _{a_1}\right|
Y\right)\  n_{1}\left( \left. \rho _{0}\rho _{1},\rho _{b_0}\rho _{b_1}\right|
Y\right) ,  \tag{14}
\end{eqnarray}
where the normalisation factor takes into account the
overcompleteness of the eigenbasis, since $E^{-n,-\nu}$ and $E^{n,\nu}$ correspond to the same component $n_{2}^{n,\nu}$.

Introducing finally the components $n_{1}^{n,\nu},$ see (2), for the single density
distributions
 with their corresponding  eigenvalues $\omega \left( n_{a},\nu_{a}\right) ,\omega \left(
n_{b},\nu_{b}\right) $ and  eigenvectors $E^{n_{a},\nu_{a}}{\left( \rho
_{a_0}\alpha ,\rho _{a_1\alpha }\right) },E^{n_{b},\nu_{b}}{\left( \rho
_{b_0\beta }\rho _{b_1\beta }\right) },$ one writes 
\[
n_{2}^{n,\nu}\left( \left. \rho _{\gamma };\rho _{a_0}\rho _{a_1},\rho _{b_0}\rho
_{b_1}\right| \omega \right) =\frac{1}{2a\left(n,\nu\right) \left( \omega
-\omega {\left( n,\nu\right) }\right) \left| \rho _{a}\rho _{b}\right| ^{2}}\times %
\]
\[
\,\,\times \stackunder{na,nb}{\sum }\int\frac{d\nu_{a}d\nu_{b}}{a\left(n_{a},\nu_{a}\right) a\left(n_{b},\nu_{b}\right) }\ \frac{1}{\omega {\left(
n_{a},\nu_{a}\right) }+\omega {\left( n_{b},\nu_{b}\right) }-\omega {\left(
n,\nu\right) }} 
\]
\[
\times \int d^{2}\rho _\alpha d^{2}\rho _\beta \ \bar{E}^{n_{a},\nu_{a}}{\left( \rho
_{a_0 \alpha} ,\rho _{a_1\alpha} \right) }\bar{E}^{n_{b},\nu_{b}}{\left(
\rho _{b_0\beta} ,\rho _{b_1\beta} \right) } \qquad \qquad
\qquad \,\qquad 
\]
\begin{equation}
\times \int \frac{d^{2}\rho _{0}d^{2}\rho _{1}d^{2}\rho _{2}}{\left| \rho
_{01}\ \rho _{02}\ \rho _{12}\right| ^{2}}\ E^{n,\nu}{\left( \rho _{0\gamma },\rho
_{1\gamma }\right) }E^{n_{a},\nu_{a}}{\left( \rho _{0\alpha },\rho _{2\alpha }\right)
}E^{n_{b},\nu_{b}}{\left( \rho _{1\beta} ,\rho _{2\beta} \right)
}\ ,\,\qquad \,\,\,\,\qquad  \tag{15}
\end{equation}
where $\rho _{a}=\rho _{a_0}\!-\!\rho _{a_1},\ \rho _{b}=\rho _{b_0}\!-\!\rho _{b_1},$ and $\rho _{\alpha },\rho _{\beta }$ are auxiliary  coordinates which
play for the dipoles $\left( \rho _{0}\rho _{2}\right) $ and $\left( \rho
_{1}\rho _{2}\right) $  the same r\^ ole played by $\rho _{\gamma 
\text{ }}$ for the dipole $\left( \rho _{0}\rho _{1}\right) $, see Fig.1. The
expression (15) gives the formal solution of $n_{2}$ in terms of the
eigenfunctions of the BFKL kernel which are given in (4).

Let us  comment one after the other the 3 different building blocks appearing in  (15)

i)\quad The summation ${\sum_{na,nb} }\int d\nu_{a}d\nu_{b}$
is related to the  dipoles $\left( \rho _{a_0}\rho
_{a_1}\right) , $ $\left( \rho _{b_0}\rho _{b_1}\right) $ and will be used
for inserting $n_{2}$ in cross-section calculations. It will
also restrict the final integration over the quantum numbers $\left(
n,\nu\right) $ through the poles appearing at $\omega = \omega \left(
n,\nu\right) =\omega \left( n_{a},\nu_{a}\right) +\omega \left(
n_{b},\nu_{b}\right) $ in the  denominators of (15).

ii)\quad The factors $\bar{E}^{n_{a},\nu_{a}}{\left( \rho _{a_0\alpha} ,\rho _{a_1\alpha
}\right) }\bar{E}^{n_{b},\nu_{b}}{\left( \rho _{b_0\beta }\rho _{b_1\beta
}\right) }$ are the only ones depending on the coordinates
 $\left( \rho _{a_0}\rho _{a_1}\right) $ and $\left( \rho _{b_0}\rho
_{b_1}\right) $. Interestingly enough, the subsequent interaction terms
involving these dipoles, for instance when computing multi-Pomeron
interactions in the QCD dipole picture\cite {muller}, will be greatly simplified by using  factorization and
the orthogonality relations (11). These features are already at the root of the
derivation\cite {Navelet}
 of the equivalence (see formula (1)) between the elastic BFKL and QCD-dipole amplitudes.

iii)\quad The last factor in the solution (15), is the
triple-dipole vertex  $V_{\alpha \beta \gamma} $
 in the QCD dipole picture. It depends on the conformal quantum numbers and auxiliary coordinates of the 3 dipoles which are coupled together in $n_{2}.$ It is the purpose
of the next sections to give an interpretation of these
vertices in terms of dual amplitudes.

\section{The triple-dipole vertex as a dual Shapiro-Virasoro amplitude}

Let us first consider the vertex $V$ for
conformal spins $n_{a}\!=\!n_{b}\!=\!n\!=0.$ This vertex  is physically relevant
since it corresponds to the dominant contribution at high $Y.$%
 In fact, the method used for
 its evaluation will be valid for any conformal spin.
Inserting the definitions (4) in the expression of $V_{\alpha \beta \gamma } 
$ obtained from the solution (15), one gets: 
\begin{equation}
V_{\alpha \beta \gamma }=\stackunder{{\Bbb C}^{3}}{\int }\frac{d^{2}\rho
_{0}d^{2}\rho _{1}d^{2}\rho _{2}}{\left| \rho _{01}\!\rho _{02}\!\rho
_{12}\right| ^{2}}\ \left| \frac{\rho _{0\gamma }\rho _{1\gamma }}{\rho
_{01}}\right| ^{-2i\nu-1}\left| \frac{\rho _{1\beta }\rho _{2\beta }}{\rho
_{12}}\right| ^{-2i\nu_{b}-1}\left| \frac{\rho _{0\alpha }\rho _{2\alpha }}{%
\rho _{02}}\right| ^{-2i\nu_{a}-1}.  \tag{16}
\end{equation}
Our observation is that \medskip $V_{\alpha \beta \gamma }$ can be
expressed as follows : 
\begin{equation}
V_{\alpha \beta \gamma }=\nu_{\alpha \beta \gamma }^{\nu ,\nu _{a},\nu
_{b}}\ B_{6}\left( \nu ,\nu _{a},\nu _{b}\right) ,  \tag{17}
\end{equation}
where $B_{6}$ is a Shapiro Virasoro\cite{Shapiro-Virasoro} amplitude with $6$ external legs and $%
\nu^{\nu ,\nu _{a},\nu _{b}}_{ \alpha \beta\gamma }$ is a known conformally-invariant tensor of the coordinates, i.e. it is completely fixed\cite{Polyakov} by the
symmetry. In order to derive the expression (17), let us recall the Koba-Nielsen formulation\cite{Koba-Nielsen} of $%
B_{6}$, namely. 
\begin{equation}
B_{6}=\int \frac{d^{2}\rho _{0}d^{2}\rho _{1}d^{2}\rho _{2}}{\left| \rho _{\alpha
\beta }\ \rho _{\beta \gamma }\ \rho _{\gamma \alpha }\right| ^{-2}} 
\stackunder{\stackunder{i,j=\left\{ 0,1,2,\alpha ,\beta ,\gamma \right\} }{%
i<j}}{\sum }\left| \rho _{ij}\right| ^{-2p_{ij}},  \tag{18}
\end{equation}
where the integration measure is the conformally-invariant one and
the powers $p_{ij}$ correspond to scalar products of external momenta on
the target space in the closed string realizations of the Shapiro-Virasoro amplitudes. Note that the string tension can be chosen to be equal to 1 by scale invariance of the solution. In our case, the  powers $p_{ij}$ are not specified in terms of  external momenta but stringent constraints are required on $p_{ij}$ in order to
satisfy the requirements of duality and conformal symmetry\cite{Frampton}
namely: 
\begin{equation}
\forall _{i}: \ \ p_{ii}=-2\ \qquad \stackunder{j=1}{\stackrel{6}{\sum }}p_{ij}=0;
\tag{19}
\end{equation}
It is tedious but straightforward   to verify that the constraints (19)
are verified for the set of  $p_{ij}$ corresponding to  formula (16) given in Table I. From that, 
one easily obtains formula (17) with 
\begin{equation}
\nu_{\alpha \beta \gamma }^{\nu ,\nu _{a},\nu _{b}}=\left| \rho _{\alpha \beta
}\right| ^{2i\left( \nu -\nu _{a}-\nu _{b}\right) -1}\left| \rho _{\beta
\gamma }\right| ^{2i\left( \nu _{a}-\nu _{b}-\nu \right) -1}\left| \rho_{
\alpha \gamma }\right| ^{2i\left( \nu _{b}-\nu _{a}-\nu \right) -1}.  \tag{20}
\end{equation}
Formula (20) does not come as a surprise, since it is 
 the universal form\cite{Polyakov} of a 3-point
correlation function for field theories obeying global conformal invariance. Indeed it corresponds to a correlation function  $$\left\langle {\cal O}^{\Delta }{\left( \rho _\gamma \right)
}{\cal O}^{\Delta _a}{\left( \rho _\alpha \right) }{\cal O}^{\Delta _b}{\left(
\rho _\beta \right) }\right\rangle $$ where ${\cal O}^{\Delta },$ ${\cal O}^{\Delta _a},$ ${\cal O}^{\Delta _b}$ are so-called ``quasi-primary'' fields%
\cite{Polyakov}
of conformal dimensions $\frac{1}{2}-i\nu,\frac{1}{2}-i\nu _{a},\frac{1}{2}%
-i\nu _{b},$ respectively. It is interesting to note that such fields already
appear  in the field-theoretical interpretation of the BFKL equation and its relation to conformal invariance
properties\cite{Lipatov}, where correlation functions  
\begin{equation}
\left\langle \varphi {\left( \rho _\alpha \right) }\varphi {\left( \rho_
\beta \right) }{\cal O}^{\Delta }{\left( \rho_ \gamma \right) }\right\rangle
\ \propto \  E^{n,\nu }{\left( \rho _{\alpha \gamma} ,\rho _{\beta \gamma} \right) } 
\tag{21}
\end{equation}
are introduced, with $\varphi $ being a scalar (i.e. of conformal  dimension 0) field  representing
 external (reggeized) gluons\footnote{Reggeized gluons are known to have no spin degree of freedom in the high-energy limit of QCD.}. In this sense our result (17)
can be considered as a QCD dipole model realization of the 
triple hard-Pomeron vertex. It is a stimulating (but non-trivial) challenge to
confront the solution we obtain (in particular Eqns.(15-20)) with the known results\cite {Bartels1} on the $(2\!\rightarrow \! 4)$ gluon vertex in the BFKL approach. This would allow to go deeper in the equivalence of BFKL to the QCD dipole approach\footnote{See footnote 1.}.

 The result obtained in Eqn.(20) for \medskip $\nu^{\nu ,\nu _{a},\nu _{b}}_{\alpha \beta
\gamma }$ is a consequence of the global conformal
invariance of the BFKL equation. Indeed, the $B_{6}$ function is conformally invariant by
construction (provided the constraints (19) are satisfied) and the overall
vertex $V_{\alpha \beta \gamma }$ of formula (17) should reflect the
general structure\cite{Polyakov}
\begin{equation}
\left\langle {\cal O}^{\Delta }{\left( \rho _\gamma \right) }{\cal O}%
^{\Delta a}{\left( \rho _\alpha \right) }{\cal O}^{\Delta b}{\left( \rho _\beta \right)
}\right\rangle \equiv \frac{C{\left( \Delta ,\Delta _{a},\Delta
_{b}\right) }}{\left| \rho _{\alpha \beta }\right| ^{2\left( \Delta
_{a}+\Delta _{b}-\Delta \right) }\left| \rho _{\beta \gamma }\right|
^{^{2\left( \Delta _{b}+\Delta -\Delta _{a}\right) }}\left| \rho _{\gamma
\alpha }\right| ^{^{^{2\left( \Delta _{a}+\Delta -\Delta _{b}\right) }}}}\ , 
\tag{22}
\end{equation}
where $C$ $\left( \Delta ,\Delta _{a},\Delta _{b}\right) $ has to be 
coordinate-independent while its dependence on the conformal dimensions is not specified by the global $SL\left( 2,{\Bbb C}\right) $
symmetry. This result could be expected for any conformally invariant
kernel. By contrast, the specific BFKL kernel $\left\{ {\cal L}ip\right\}$ of formula (9) allows the determination
of $C \left( \Delta ,\Delta _{a},\Delta _{b}\right) $ and its relation to the
Shapiro-Virasoro amplitude $B_{6}$.

\section{$(1\!\rightarrow \!p)$ dipole vertex}

Let us generalize the vertex calculation to the $(1\!\rightarrow \!p)$-dipole
amplitude by considering first the exemple of $n_{3}^{n,\nu }\left( \left. \rho
_{0}\rho _{1};\rho _{a_0}\rho _{a_1},\rho _{b_0}\rho _{b_1},\rho _{c_0}\rho
_{c_1}\right| {\omega }\right) ,$ see  Fig.2. In much the same way as for $%
n_{2}^{n,\nu },$ one writes: 
\[
n_{3}^{n,\nu }\left( \left. \rho _{\gamma };\rho _{a_0}\rho _{a_1},\rho
_{b_0}\rho _{b_1},\rho _{c_0}\rho _{c_1}\right| {\omega }\right) =\frac{1}{%
2a\left( n,\nu \right) \left( \omega -\omega {\left( n,\nu \right) }\right) 
}\frac{1}{\left| \rho _{a}\ \rho _{b}\ \rho _{c}\right| }\times 
\]
\begin{eqnarray}
&& \!\!\!\!\!\!\!\!\sum_{n_{a},n_{b},n_{c}}\int \frac{d\nu _{a}d\nu _{b}d\nu
_{c}}{a{\left( n_{a},\nu _{a}\right) }a{\left( n_{b},\nu _{b}\right)
}a{\left( n_{c},\nu c\right) }}\frac{1}{\omega {\left( n_{a},\nu
_{a}\right) }+\omega {\left( n_{b},\nu _{b}\right) }+\omega {\left(
n_{c},\nu _{c}\right) }-\omega}   \nonumber \\
 \!\!\!\!\!\!\!\! &&\int d^{2}\rho _{\alpha }d^{2}\rho _{\beta }d^{2}\rho
_{\delta }
\;\bar{E}^{n_{a},\nu _{a}}{\left( \rho _{a_{0}\alpha},\rho _{a_{1}\alpha}
\right)}
\;\bar{E}^{n_{b},\nu _{b}}{\left( \rho _{b_{0}\beta},\rho _{b_{1}\beta}
\right)}
\;\bar{E}^{n_{c},\nu _{c}}{\left( \rho _{c_{0}\delta},\rho _{c_{1}\delta}
\right)}\times
  \nonumber \\
\!\!\!\!\!\!\!\! &&\int {\scriptstyle \frac{d^{2}\rho _{0}d^{2}\rho _{1}d^{2}\rho _{2}d^{2}\rho _{3}%
}{\left| \rho _{01}\ \rho _{12}\ \rho _{23}\ \rho _{30}\right| ^{2}}\ E^{n,\nu }\left( \rho
_{0\gamma },\rho _{1\gamma }\right) E^{n_{a},\nu _{a}}\left( \rho _{1\alpha
},\rho _{2\alpha }\right) E^{n_{b},\nu _{b}}\left( \rho _{2\beta },\rho
_{3\beta }\right) E^{n_{c},\nu _{c}}\left( \rho _{3\delta },\rho
_{0\delta }\right)},  \tag{23}
\end{eqnarray}
\noindent where we have used the fact that the probability of finding three
dipoles at  $Y$ can be expressed by  two equivalent iterations of the BFKL kernel, namely:
\begin{equation}
\left|\frac {\rho _{01}}{ \rho _{03}\ \rho _{13}}\right|^2\times
\left|\frac {\rho _{13}}{ \rho _{32}\ \rho _{21}}\right|^2
\equiv
\left|\frac {\rho _{01}}{ \rho _{02}\ \rho _{12}}\right|^2\times
\left|\frac {\rho _{02}}{ \rho _{32}\ \rho _{30}}\right|^2
=
\left|\frac {\rho _{01}}{ \rho _{12}\ \rho _{23}\ \rho _{30} }\right|^2.
\tag{24}
\end{equation}
Hence, the intermediate step (the ${\rho _{0}}{ \rho _{2}}$ segment or, equivalently, the 
${\rho _{1}}{ \rho _{3}}$ segment
in the case of Fig.2)
is not relevant in the final integration kernel.  From the symmetry of (24), it is easy to realize  that this iteration procedure
will give a symmetric kernel for any number $p$ of produced dipoles.

The integral over the four eigenfunctions appearing in the last term of
expression (23) gives the $1\!\rightarrow \! 3$ dipole vertex and can be
cast in the following form (for  zero conformal
weights):
\begin{equation}
V_{\left\{ \alpha _{i}\right\} }
=\prod_{i<j}^{4}\ \left|\rho _{\alpha _{i}\alpha _{j}}^2\right|^{\frac23 \left[(\sum\Delta)-\Delta
_{i}-\Delta _{j}\right]}\times \ C_{8}\left( \frac{\rho _{\alpha \beta_{} }\rho _{\gamma \delta_{}
}}{\rho _{\alpha \gamma }\rho _{\beta \gamma }},\frac{\rho _{\alpha \beta
}\rho _{\gamma \delta }}{\rho _{\alpha \delta }\rho _{\beta \gamma }}\right)
,  \tag{25}
\end{equation}
\noindent where the first factor is determined\cite {Polyakov,difrancesco} by conformal invariance
constraints on 4-point correlation functions for quasi-primary fields of conformal dimension $\Delta_i.$\ It is a
 matter of tedious but straightforward verification that the vertex term in equation (23) can be
cast into the generic form (25) with:
\begin{equation}
\Delta =\frac{1}{2}-i\nu ,\;\Delta _{a}=\frac{1}{2}-i\nu _{a},\;\Delta _{b}=%
\frac{1}{2}-i\nu _{b},\;\Delta _{c}=\frac{1}{2}-i\nu _{c}.  \tag{26}
\end{equation}
\noindent Indeed the functional dependence (25) is obtained from the
constrained relations $p_{ii}=-2,\;\sum\limits_{j=1}^{8}p_{ij}=0,$ see (19). The corresponding values of  $p_{ij}$ are displayed in table II.

As well-known\cite{Polyakov} global conformal invariance of 4-point correlation functions
does not determine the residual function $C_{8},$ as an arbitrary function of the $SL\left( 2,{\Bbb C}\right) $-invariant harmonic ratios of the 8
 coordinates involved in the problem, see Fig.2. However the explicit integrand (24)
induced by  the BFKL kernels leads to a well-defined connection with the
(Koba-Nielsen form of) the Shapiro-Virasoro $B_{8}$ function. Indeed, one may write:
\begin{equation}
C_{8}=\int \prod_{i=1}^{8}d\rho _{i}\left[ \frac{d\rho _{\alpha }d\rho
_{\beta }d\rho _{\gamma }d\rho _{\delta }}{\left| \rho _{\alpha \beta }\rho
_{\beta \gamma }\rho _{\gamma \delta }\rho _{\delta \alpha }\right| ^{2}}%
\right] ^{-1}\prod_{i<j}\rho _{ij}^{-2p_{ij}},  \tag{26}
\end{equation}
\noindent where the factor $\left[ ...\right] ^{-1}$  conventionnally means 
that 4 integration variables out of $8$ have to be omitted. Formula
(26) can be identified by comparison with  the Koba-Nielsen formulation of $B_{8},$
 namely:
\begin{equation}
B_{8}=\int \prod\limits_{i=1}^{8}d\rho _{i}\left[ \frac{d\rho
_{\alpha }d\rho _{\beta }d\rho _{\delta }}{\left| \rho _{\alpha \beta }\rho
_{\beta \delta }\rho _{\delta \alpha }\right| ^{2}}\right]
^{-1}\prod\limits_{i<j}\rho _{ij}^{-2p_{ij}}\ \equiv \int d\rho _{\delta}
\left|  \frac{\rho
_{\beta \delta }}{\rho _{\gamma \delta }\rho _{\delta \alpha }} \right| ^{2}
\  C_{8},  \tag{27}
\end{equation}
\noindent where the three coordinates in the $SL\left( 2,{\Bbb C}\right) $ volume factor have been choosen in order to match with the variables of $C_8.$
 It is clear from formulae (26,27), that $C_{8}$ is  the $%
SL\left( 2,{\Bbb C}\right) $-invariant integrand of the Shapiro--Vivasoro
amplitude $B_{8}.$

By simple iteration  in the number of produced dipoles, it
is not difficult to find the generalization of formulae (23)-(27)
to the $(1\!\rightarrow \! p)$ dipole distribution. The key observation to
determine the powers $p_{ij}$ appearing in (26) --which are known combinations of conformal dimensions $\Delta_i$ obeying the
constraints (19) -- is to notice that the coordinates in transverse plane are  nearest neighbours along the polygonal perimeter $\rho_0,\rho_1,...,\rho_p,$ (see Fig.2 for the case $p=3$).
 The auxiliary points $\rho
_{\alpha _i}$ are also connected via nearest neighbours to the previously mentioned
perimeter, except for their conformal-dimension factors $p_{\alpha i,\alpha j}$ which are fixed by the constraints. Indeed, we note that the 4 legs connected to a given coordinate $\rho _{i}$ in the polygonal perimeter are such that the corresponding sum $%
\sum p_{ij}\,$ is always zero. The powers attached to the
auxiliary points are fully determined by $SL\left( 2,{\Bbb C}\right) $
invariance (for the set of constraints (19)).

\section{\protect\medskip A closed string theory for the QCD dipole picture?
}

The Shapiro-Virasoro amplitudes which are obtained for  $(1\!\rightarrow \! p)$ dipole distributions lead naturally to  the question of
a closed-string interpretation of the high-energy limit of perturbative QCD. Indeed, these amplitudes appear in the  context of a closed string moving in a Minkowskian $(1,d\!-\!1)$ target space\cite {Frampton}. More generally, such amplitudes appear as a consequence of vertex operator constructions in conformal field theories\cite {difrancesco} and are related to the existence of an (anomalous) infinite-dimensional algebra associated with local conformal invariance, namely the Virasoro algebra. Moreover in the case of a critical target-space dimension $d_c,$ the Fock Space on which the quantum string theory is realized is spanned by positive normed states (no ghosts) with full reparametrization invariance. This connection has both a practical and conceptual interest for QCD calculations. First, the many and much explored mathematical properties of dual amplitudes may lead to a simplification of QCD dipole computations for given processes, e.g. ``hard'' diffraction, multi-Pomeron contributions, etc.. Second, there is a possibility of building an effective theory of QCD in the high-enrgy limit, which could be based on a string theory (instead of a field theory). This would allow the computation of string loop contributions and thus induce an effective theory of interacting QCD Pomerons. 

However, the variables which appear as conformal exponents $p_{ij}$
of the integrands are not directly expressed as scalar products of momenta in a 
Minkowskian $(1,d\!-\!1)$ target space. They are complex numbers, see, e.g. (26) obeying constraints which are not directly expressed as on-mass shell and momentum conservation constraints as for the closed string\cite {Frampton}. Even if such a target-space interpretation is possible, an analytic continuation in the imaginary direction (implied by the quantum numbers of the conformal eigenvectors (4)) is to be performed. It is thus useful to pass in review the  properties of Shapiro-Virasoro amplitudes in this context and to see which are those to be completed for a full closed string theory to be valid.

\quad i)\quad {\bf Duality}

 A first consequence of the solutions (17), (25), and
their iterations, is that duality properties exist in the dipole formulation of QCD vertices. Indeed, by construction, Shapiro-Virasoro amplitudes are meromorphic but forbid the existence of multiple pole singularities coming from  dual channels. For instance,
in Fig.2, the $(\rho_0,\rho_2)$ and $(\rho_1,\rho_3)$  channels cannot both together bring  singularities to the amplitude. As usual in dual theories, there exists intricate relations between
different ways of describing the amplitudes depending on the series of 
pole contrbutions which are choosen for their expansion. An interesting example of such a duality property has been provided by the equivalence of the ``t-channel'' BFKL elastic 4-gluon amplitude with the ``s-channel''
QCD dipole description of the same amplitude\cite {Navelet}. Further application of this fruitful concept are expected from our results.

\quad ii)\quad {\bf Non-zero conformal spins}
 
As a practical consequence of our identification of the
multiple-dipole vertices with  integrands of standard Shapiro-Virasoro amplitudes in the case of zero conformal spins, one may use some  tools which are developed in the string theoretical formalism
to generalize our investigations to the case of general (integer or
half-integer) conformal spins as follows; One can consider in general amplitudes of the form:
\begin{equation}
B_{N}=\int \prod\limits_{i=1}^{N}d\rho _{i}\left[ \frac{d\rho
_{\alpha }d\rho _{\beta }d\rho _{\delta }}{\left| \rho _{\alpha \beta }\rho
_{\beta \delta }\rho _{\delta \alpha }\right| ^{2}}\right]
^{-1}\prod\limits_{i<j}^N
\ \rho _{ij}^{-p_{ij} + \frac {n_{ij}}2}
\bar {\rho} _{ij}^{-p_{ij} + \frac {{\tilde n}_{ij}}2},  \tag{28}
\end{equation}
\noindent where $n_{ij},\ {\tilde n}_{ij} $ are integers\footnote{%
The possibility of extending the calculation of Ref. \cite {kawai} to
amplitudes with half-integer conformal-spin has been noticed and indicated
to me by H. Navelet, I thank him for this information prior to publication.}.
Interestingly enough, in the framework of string theory,  this corresponds to consider external excited states
of the bosonic string\cite{kawai}. Moreover, the same techniques  allow to connect
closed string to open string tree amplitudes which may allow to extend to the
multiple-vertex calculations the conformal-block structure initially identified in the BFKL 4-point amplitudes\cite{Navelet}.

\quad iii)\quad {\bf Extended conformal symmetry and Virasoro algebra} 

In the seminal paper of Ref.\cite {Lipatov}, it has been noticed that it was not straightforward to extend the (already beautiful) global conformal symmetry $SL(2,{\Bbb C})$ to the infinitely dimensional conformal group in 2 dimensions. In other words, only the 6 generators ${\Bbb L}_{-1},$ ${\Bbb L}_{1},$ ${\Bbb L}_{0}$ (3 holomorphic and 3 non-holomorphic) of the Virasoro algebra were expected to generate the symmetry algebra of the BFKL kernel. The results we obtain indicate that the algebra can probably be extended to the infinite series
of locally conformal generators, i.e. the whole Virasoro algebra, at least in the QCD dipole representation\footnote {see footnote 1.}. as usual, the symmetry is expected to be anomalous due to the possibility of  a central charge\cite {Frampton,difrancesco} (conformal anomaly) at the quantum level of consistency. This issue will depend on the interpretation of a suitable $p-$independent target-space representation of the exponents $p_{ij} ($ see, for $p=2,3$, Table I and II). For instance, If an embedding  $p_{ij}\!\rightarrow \!p_i\cdot p_j$ in a Minkowskian   $(1,d\!-\!1)$ space is allowed, this will determine the central charge to be related to $d$ and the critical dimension to be $d_c = 26,$ by compensation of the ghost contribution due to  reparametrization symmetry\cite {Frampton}. This interesting issue  deserves certainly more study. 

\bigskip
{\bf Acknowledgments}
We want to  thank Andrzej Bialas and Henri Navelet for a fruitful collaboration on the QCD theory of dipoles which initiated the present work  helping the author  to find the way out in two noticeable occasions.  Christophe Royon and Samuel Wallon are acknowledged for stimulating discussions.

\medskip

\medskip

\end{document}